\documentclass{appolb}
\usepackage{epsfig}

\def\ETJ{\ensuremath{E_T^{\mathrm{jet}}}}

\newcommand {\pom} {I\!\!P}
\newcommand {\pomsub} {{\scriptscriptstyle \pom}}

\newcommand {\xpom} {x_{\pomsub}}
\newcommand {\apom} {\alpha_{\pomsub}}

\newcommand {\aprime} {\alpha^\prime_\pomsub}

\newcommand{\gp}{\gamma p}
\newcommand{\gvp}{\gamma^* p}
\newcommand{\gv}{\gamma^*}
\newcommand\units{\,\mathrm}

\newcommand{\gevtwo}{\units{GeV^2}}
\newcommand{\gevmtwo}{\units{GeV^{-2}}}

\newcommand{\ftwod}{F_2^D}

\newcommand{\ftwopom}{F_2^{\pomsub}}

\def\ETJ{E_T^{\mathrm{jet}}}
\def\ETJ1{E_T^{\mathrm{jet1}}}
\def\ETAJ{\eta^{\mathrm{jet}}}

\begin{document}
\title{The HERA challenges for LHC%
\thanks{Presented at the 2009 Epiphany meeting, Cracow}%
}
\author{Halina Abramowicz
\address{The Raymond and Beverly Sackler School of Physics and Astronomy, Tel Aviv University, 69978 Tel Aviv, Israel}
}
\maketitle
\begin{abstract}
  Over the last two decades, the HERA collider has provided a large
  amount of new information about QCD dynamics at high energy. While
  the most appreciated are the measurements of the proton structure
  functions in a wide range of parton momentum $x$ and virtuality
  $Q^2$, it is hard to believe that some of the observations at HERA
  which do not fit the simple picture of DGLAP dynamics would not get
  amplified at the LHC, possibly rendering certain approaches to
  searches beyond the Standard Model inadequate.
\end{abstract}
\PACS{12.38.Aw, 12.38.Bx, 12.39.St, 13.60.-r, 13.85.-t,13.87.-a}
  
\section{Introduction}

Progress in the next decade (and most probably even longer) in High
Energy Physics will be totally overshadowed by the results expected
from the Large Hadron Collider (LHC) at CERN (Geneva) where high
energy protons will collide at an unprecedented center of mass energy
of 14 TeV, any time soon. For quite some time, theorists claim that
the success of the Standard Model (SM) of the unified electroweak
interactions and of strong interactions cannot be understood, unless
it is a low energy realization of a more sophisticated construct,
which should manifest itself already at TeV scales.  It still remains
to be seen whether the Standard Model Higgs, whose mass is constrained
by precision electroweak measurements~\cite{higgsmass}, is realized in
nature. A low mass of the SM Higgs, about 120 GeV, creates the
so-called hierarchy problem, that is the need at high energy for a
tremendous fine-tuning of quantum-loop corrections to keep the Higgs
mass light enough for it to be relevant for electroweak symmetry
breaking. The favored candidates for the new horizon are
supersymmetric (SUSY) theories and/or theories with universal
extra-dimensions (UED).  There are many realizations of these theories
(see~\cite{jellis} for a recent overview).  Their common denominator
is the appearance of new particles or states, which will be signaled
by a departure of the measured cross sections from expectations of the
SM.

The generic LHC signatures of SUSY models are isolated leptons, few
energetic jets and, for $R$-parity conserving SUSY, large missing
transverse energy~\cite{dejong}. Similar signatures may be expected in
the various realizations of the UED theories~\cite{vacavant,datta}.
These signatures may also originate from the SM, dominated by hard QCD
interactions. In the context of LHC physics, QCD constitutes the
background (though for many it is in itself a interesting laboratory
for the dynamics of high energy QCD). The question is then how
reliable is the background estimate. This is where the QCD studies at
the HERA $ep$ collider become very relevant, though the energy scales
are mostly lower than the ones expected at LHC. The same could be said
of the Tevatron $p\bar{p}$ collider studies, if not for the fact that
the problems encountered at the Tevatron can only become worse at the
LHC.  And this is because of the composite nature of hadrons, nucleons
in this particular case, which have a rich dynamical internal
structure, that cannot be derived from perturbative QCD. This
structure is directly probed in the deep inelastic scattering (DIS) of
leptons on hadrons.
\section{Hard $pp$ interactions}
The reactions that will be mostly investigated at the LHC are hard
$pp$ interactions in which partons in the proton, quarks and gluons,
interact producing either high transverse momentum partons that
materialize in the form of hadronic jets or new heavy particles which
decay into leptonic and/or hadronic final states. The cross section for
the hard interactions can be calculated in perturbative QCD, though to
a limited precision due to higher order corrections. The hadronization
process is soft in nature and requires modeling. Due to the
complicated nature of resulting final states, even the expectations
from pure QCD processes require MC simulation, which limit the
precision with which the QCD background can be estimated.

There are the following four components in modeling hadronic final
states in hard $pp$ collisions~\cite{sjostrand}:\\
{\bf 1. the hard subprocess at the parton level}. At leading order it
is a $2 \rightarrow 2$ process and the cross section is expressed
through
\begin{equation}
\frac{d\sigma_{\mathrm{int}}}{dp_T^2}=\sum_{i,j,k}\int dx_1 \int dx_2 \int d\hat{t} f_i(x_1,Q^2)
f_j(x_2,Q^2)\frac{d\hat{\sigma}_{ij\rightarrow kl}}{d\hat{t}} 
\delta\left(p_T^2-\frac{\hat{t}\hat{u}}{\hat{s}}\right)\ ,
\label{eq:sigma}
\end{equation}
where $p_T$ is the transverse momentum of the outgoing partons $k$ and
$l$ and $\hat{s}=x_1 x_2 s$ is the squared center of mass energy of
the partonic scattering, with $s$ the $pp$ center of mass energy
squared and $x_{i,j}$ the fraction of the proton momenta carried by
the incoming partons $i$ and $j$. The cross section for parton
scattering is given by $\hat{\sigma}$, calculable in perturbative QCD,
and the probability of finding the partons in the proton is given by
the parton density functions (PDF) $f_i(x,Q^2)$, where the hard scale
$Q^2$ is assumed to be $Q^2=p_T^2$.  The variables $\hat{t}$,
$\hat{u}$ and $\hat{s}$ are the usual Mandelstam variables;\\
{\bf 2. initial and final state gluon radiation} from the interacting
partons, which is meant to mimic higher order QCD processes;\\
{\bf 3. additional semi-hard or hard interactions} between the
remaining partons. They arise naturally in that the integrated cross
section given by~(\ref{eq:sigma}) diverges for low $p_T$ values and
$\sigma_{\mathrm{int}}$ becomes bigger than the total cross section,
unless multi-parton interactions (MPI) are introduced;  \\ {\bf 4.
  hadronization}, the formation of hadrons from partons and from beam
remnants.

Each of these steps is subject to modeling and various Monte Carlo
generators~\cite{pythia,herwig,isajet,sherpa,alpgen} adopt different
approaches, in particular for the last three steps.  Each one of the
modeling steps is accompanied by uncertainties that are hard to
quantify. In the following, the discussion will focus on the
extraction of parton density functions which are essential for the
calculation of the cross sections and on the validity of the DGLAP
evolution in describing the QCD dynamics at high energy.
 
\section{Parton density functions in the proton}
There are many parameterizations of the proton PDFs derived in
global fits of the NLO DGLAP evolution equations to the proton structure
functions and hard scattering measurements.  The most commonly used are
the ones provided by the MRST(MSTW)~\cite{mrst} and the CTEQ
groups~\cite{cteq}. They incorporate not only the DIS data, but also
data from hadron-hadron interactions. A big effort is now vested in
quantifying the uncertainties~\cite{lhcpdf} for the precision
measurements at the LHC.  The worrisome part is that these
uncertainties usually come out smaller than the differences between
the various groups and this is particularly true with the gluon
distribution, the contribution of which dominates the QCD cross
section at LHC. 
\begin{figure}[hbt]
\begin{center}
\includegraphics[width=0.4\textwidth]{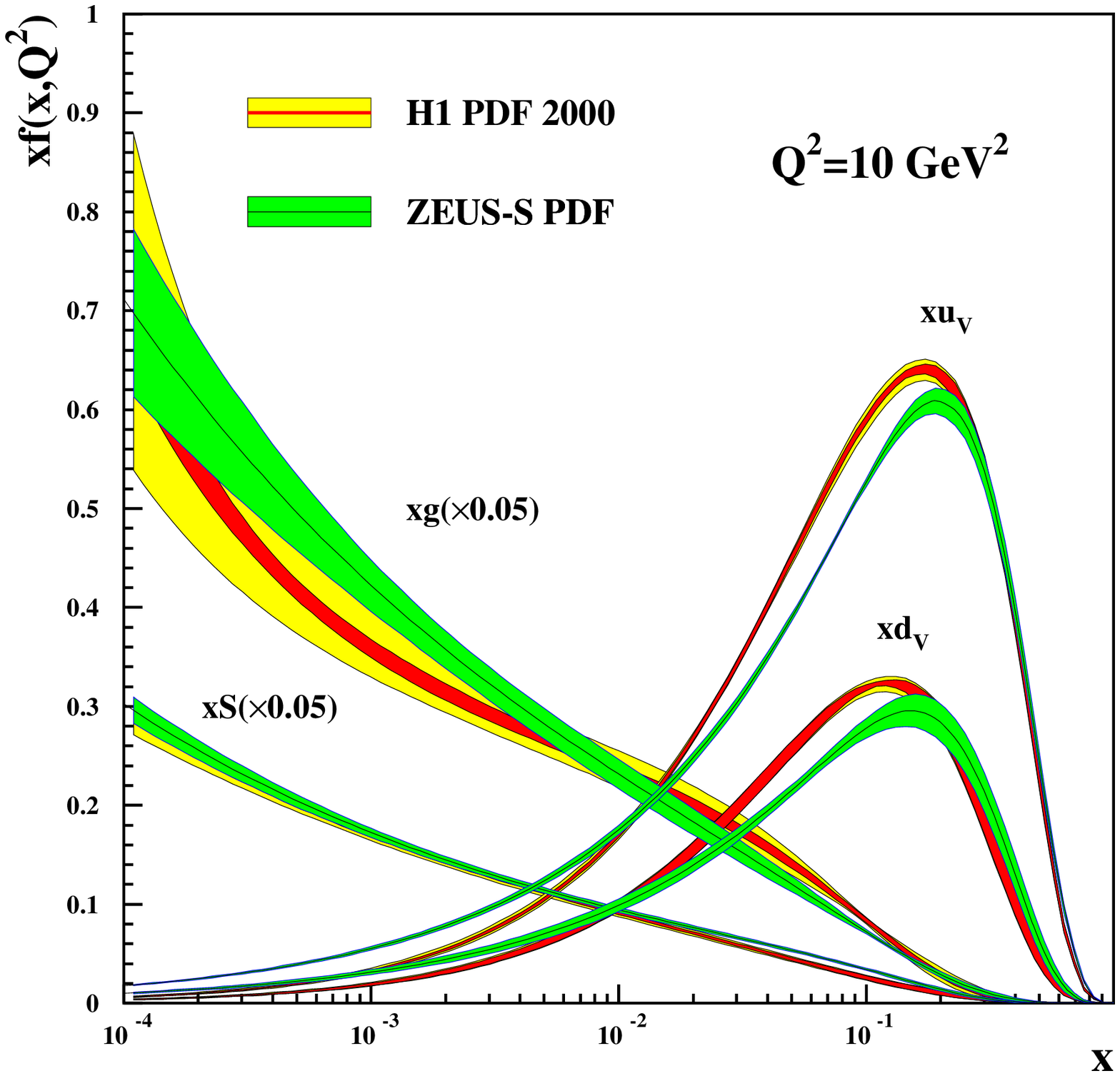}
\includegraphics[width=0.4\textwidth]{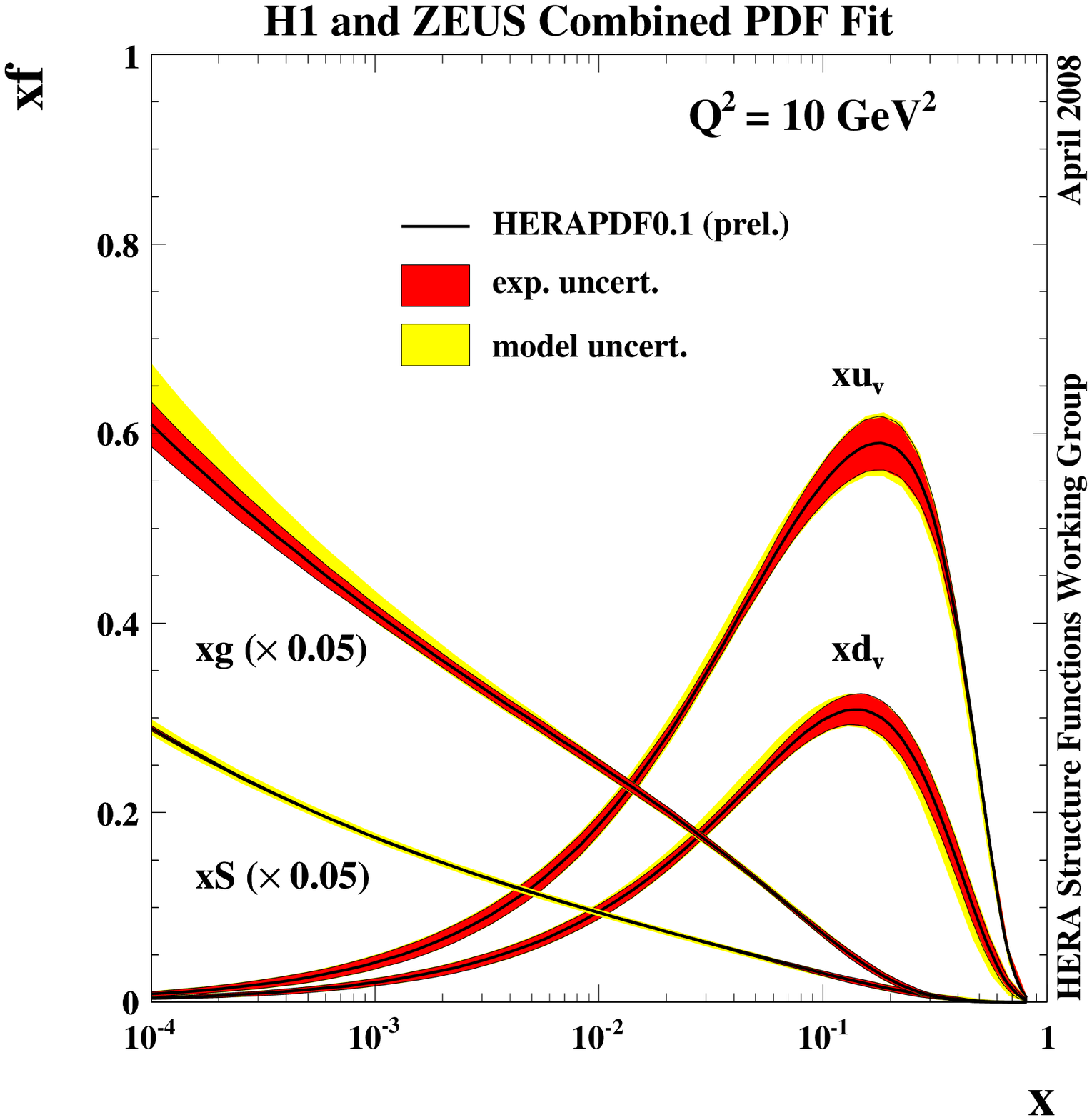}
\end{center}
\caption{Left -- comparison of the H1 and the ZEUS PDFs as a function of $x$ 
  at $Q^2=10 \gevtwo$ and right -- the result of fitting PDFs to
  the combined measurements of $F_2$ at the same $Q^2$ value.}
\label{fig:herapdf}
\end{figure}
Recently, the two HERA experiments, H1 and ZEUS, provided the most
precise PDFs by combining their HERAI measurements of the $F_2$
structure function of the proton~\cite{ichep08f2}. The
``cross-calibration'' reduces dramatically the systematic
uncertainties, as shown in Fig.~\ref{fig:herapdf}. Soon, one may
expect even better precision, when the high luminosity measurements of
neutral current and charged currents $e^\pm p$ interactions from
HERAII become available.

\subsection{Low $x$ regime}
\begin{figure}[hbt]
\begin{center}
\includegraphics[width=0.5\textwidth]{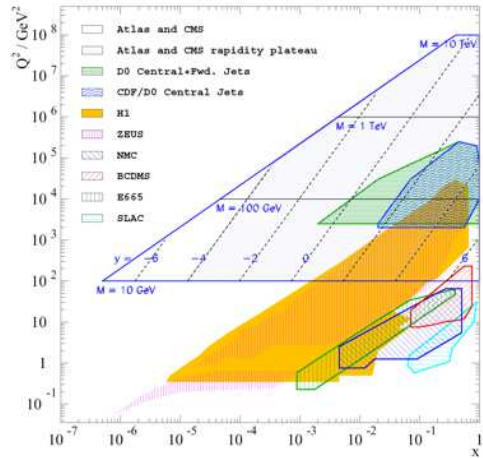}
\end{center}
\caption{The coverage of the $x$ and $Q^2$ plane by existing 
  measurements and the expected coverage in the LHC $pp$ experiments.}
\label{fig:xq2plane}
\end{figure}

\begin{figure}[hbt]
\begin{center}
\includegraphics[width=0.4\textwidth]{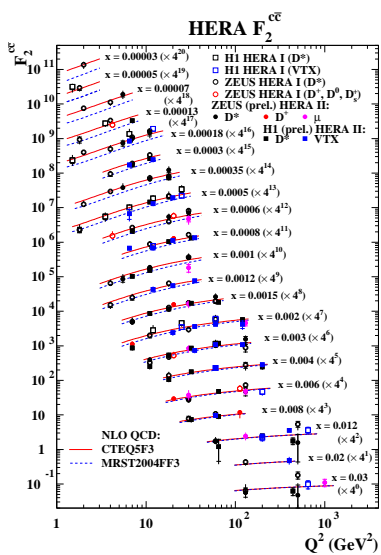}
\end{center}
\caption{Compilation of $F_2^{cc}$ as a function of $Q^2$ for different values of $x$, 
compared to expectations of two particular PDFs obtained with the fixed flavor scheme.}
\label{fig:f2cc}
\end{figure}

The extraction of PDFs assumes the validity of the DGLAP evolution
equation in the full range probed by the data, including the very low
$x$ regime of HERA. The only marker justifying this approach is the
good $\chi^2$ of the global fit. However many parameters are involved
and the lever arm at low $x$ is relatively small (see Fig.~\ref{fig:xq2plane}).
A large fraction of the low $x$ cross section in $ep$ scattering is
due to heavy flavor production, in particular $c\bar{c}$ pairs. A
compilation of HERA measurements~\cite{labarga} of $F_2^{cc}$, the
contribution of charm production to $F_2$, is shown in
Fig.~\ref{fig:f2cc}. The treatment of heavy flavor contribution to
$F_2$ is constantly evolving~\cite{thornehf} and even within the same
scheme there are large differences in the expectations of various
groups, which must affect the uncertainty on the gluon distribution.

In the DGLAP evolution equations, large gluon densities are expected
at low $x$. If the gluon density per unit area becomes large, the
probability of recombination of gluons becomes high and non-linear
effects in the evolution may show up~\cite{GLR} which would slow down
the increase of the gluon density. In addition coherent effects may be
expected~\cite{halfms}. It is quite clear by now, that the HERA data
do not have the power to rule out or confirm the presence of such
effects. However, there are indications in the HERA data that the
low $x$ regime of DIS scattering may not be fully described by the
DGLAP dynamics. This is exemplified by the large fraction of
diffractive-like events as well as by the observation of an excess of
forward jets in the low $x$ regime.

\section{Diffractive scattering in DIS}

If particle production in hard scattering is driven by the DGLAP
evolution, large rapidity gaps (LRG) between hadrons are exponentially
suppressed. Yet, at HERA, about 10\% of the DIS events have a large
rapidity gap, which separates the initial proton from the rest of the
hadronic final state~\cite{diffHERA}. It is conceivable that the
origin of these LRG is soft in nature and that their presence is
accounted for in the initial conditions of the evolution, which would
preserve the validity of the DGLAP dynamics in sufficiently hard
processes.

The presence of LRG at high energy is associated with diffractive
scattering which, in soft hadron-hadron collisions, is believed to be
due to the exchange of a Regge trajectory~\cite{regge} with vacuum
quantum numbers - the Pomeron ($\pom$). The $\pom$ trajectory is
characterized by two parameters, the intercept $\apom(0)$ and the
slope $\aprime$.  Both have been determined in soft interactions and
the most commonly cited values are $\apom(0)=1.08$~\cite{DLapom} and
$\aprime=0.250~\gevmtwo$~\cite{DLaprime}. The intercept drives the
energy dependence of the total cross section, $\sigma_{\mathrm{tot}}
\propto s^{\apom(0)-1}$. The value of $\apom$ was later
updated~\cite{cudell} to $\apom(0)=1.096^{+0.012}_{-0.009}$.

\begin{figure}[hbt]
\begin{center}
\includegraphics[width=0.4\textwidth]{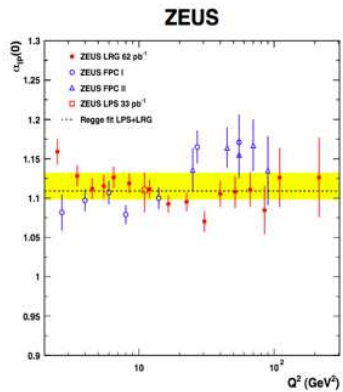}
\end{center}
\caption{The value of $\apom(0)$ extracted from diffractive DIS
  scattering as a function of $Q^2$.The dashed line with the yellow
  band represents the average value with its uncertainty.}
\label{fig:apom}
\end{figure}

The measurements of the diffractive cross section in DIS can be used
to extract the value of $\apom(0)$~\cite{h1diff,zeusdiff}. A
compilation based on the ZEUS measurements~\cite{zeusdiff} is
presented in Fig.~\ref{fig:apom}. 
The extracted value of $\apom(0)$ is shown as a function of $Q^2$ and
no dependence on $Q^2$ is observed. The averaged value of $\apom(0)$
in DIS tends to lie above the one derived from soft hadron-hadron
interactions. A larger value of $\apom(0)$ would imply a faster growth
of the cross section with energy and diffractive scattering in DIS
a harder process than in hadron-hadron scattering.

\section{QCD factorization in diffraction}

The diffractive contribution to the inclusive structure function $F_2$
is called $F_2^D$.  In addition to the usual DIS variables, $F_2^D$
also depends on variables which describe the diffractive final state,
\begin{eqnarray}
t &=& (p-p^\prime)^2 \, ,
\label{eq:deft} \\
\xpom &=& \frac{q\cdot (p-p^\prime)}{q \cdot p} \, ,
\label{eq:defxpom} \\
\beta &=& \frac{Q^2}{2q \cdot (p-p^\prime)} = \frac{x}{\xpom}\, .
\label{eq:defbeta}
\end{eqnarray}
Here $p$ and $p^\prime$ respectively denote the proton four-vector
before and after the scattering, $q$ is the four-vector of the virtual
photon $\gv$ with $Q^2=-q^2$.  The variable $t$ is the square of the
four-momentum exchanged in the proton vertex. The variable $\xpom$ is
the fractional proton momentum which participates in the interaction
with $\gv$ (sometimes denoted by $\xi$), $\beta$ is the equivalent of
Bjorken $x$ but relative to the exchanged object and $\xpom\cdot\beta
= x$.  QCD factorization is expected to hold for
$\ftwod$~\cite{Collins:1998,Berera:1996,Trentadue:1994} and
it may then be decomposed into diffractive parton distributions (DPDF),
$f^D_i$, in a way similar to the inclusive $F_2$,
\begin{equation}
\frac{d\ftwod (x,Q^2,\xpom,t)}{d\xpom dt}=\sum_i \int_0^{\xpom} dz 
\frac{d f^D_i(z,\mu,\xpom,t)}{d\xpom dt} \hat{F}_{2,i}(\frac{x}{z},Q^2,\mu) \, ,
\label{eq:fact}
\end{equation}
where $\hat{F}_{2,i}$ is the universal structure function for DIS on
parton $i$, $\mu$ is the factorization scale at which $f^D_i$ are
probed and $z$ is the fraction of momentum of the proton carried by
the diffractive parton $i$. Diffractive partons are to be understood
as those partons in the proton from which the scattering leads to a
diffractive final state. The DGLAP evolution equation applies in the
same way as for the inclusive case.  For a fixed value of $\xpom$, the
evolution in $x$ and $Q^2$ is equivalent to the evolution in $\beta$
and $Q^2$.

If, following Ingelman and Schlein~\cite{Ingelman:1985}, one further
assumes the validity of Regge factorization, $\ftwod$ may be decomposed
into a universal $\pom$ flux and the structure function of the $\pom$,
\begin{equation}
\frac{d\ftwod (x,Q^2,\xpom,t)}{d\xpom dt}= f_{\pom/p}(\xpom,t)
\ftwopom (\beta,Q^2) \, ,
\label{eq:f2pom}
\end{equation}
where the normalization of either of the two components is arbitrary.
It implies that the $\xpom$ and $t$ dependence of the diffractive
cross section is universal, independent of $Q^2$ and $\beta$, and given
by
\begin{equation}
f_{\pom/p}(\xpom,t) \sim \left( \frac{1}{\xpom} \right)^{2\apom(0)-1}
e^{(b_0^{D}-2\aprime \ln \xpom)t} \, ,
\label{eq:pomflux}
\end{equation}
one of the expectations which agrees with the measurements as
discussed in the previous section.

In this approach, the mechanism for producing LRG is assumed to be
present at some scale and the evolution formalism allows to probe the
underlying partonic structure. The latter depends on the coupling of
quarks and gluons to the Pomeron.

\begin{figure}[hbt]
\begin{center}
\includegraphics[width=0.4\textwidth]{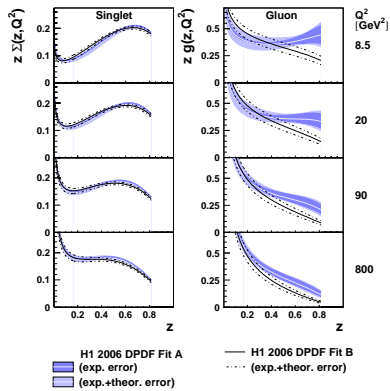}
\end{center}
\caption{The singlet, $\Sigma$, and the gluon distribution in
  diffractive parton distributions as a function of the fraction $z$
  of the momentum of the colorless exchange responsible for the LRG
  and for different $Q^2$ values. Two different fit results are
  presented as described in the figure.}
\label{fig:dpdf}
\end{figure}

An example of diffractive parton distributions derived by the H1
experiment is shown in Fig.~\ref{fig:dpdf}~\cite{h1diff}.
As in the case of the proton PDF, the gluons are not well constrained
by the inclusive measurements. They are further constrained by
diffractive jet~\cite{h1diffjets} and heavy
flavor~\cite{zeusdiffcc,h1diffcc} production. The fact that all
diffractive DIS data can be accommodated by the same set of DPDF lands
support to QCD factorization.

Gluons constitute a large fraction (about 70\%) of the diffractive
exchange, implying that DIS induced by gluons is more likely to lead to
a diffractive final state. In the black-body limit the fraction of
diffractive events cannot exceed 50\%~\cite{pumplin,kaidalov}. At HERA, this
limit is far from being reached, however the dominance of gluons in
diffractive scattering indicates that the onset of unitarity effects
may first show up in the gluon.

\section{QCD factorization breaking in diffractive scattering}

\begin{figure}[hbt]
\begin{center}
\includegraphics[width=0.4\textwidth]{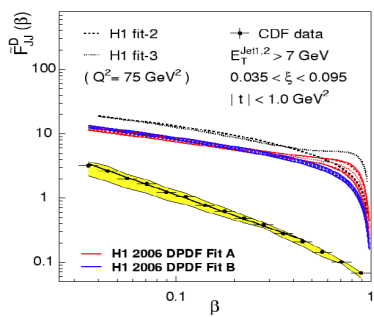}
\end{center}
\caption{The ratio of diffractively produced dijets to inclusive
  dijets as a function of $\beta$ in $p\bar{p}$ collisions compared
  to expectations of various DPDFs, as indicated in the figure,
  extracted from the HERA data.}
\label{fig:cdfdiff}
\end{figure}

\begin{figure}[hbt]
\begin{center}
\includegraphics[height=0.13\textheight]{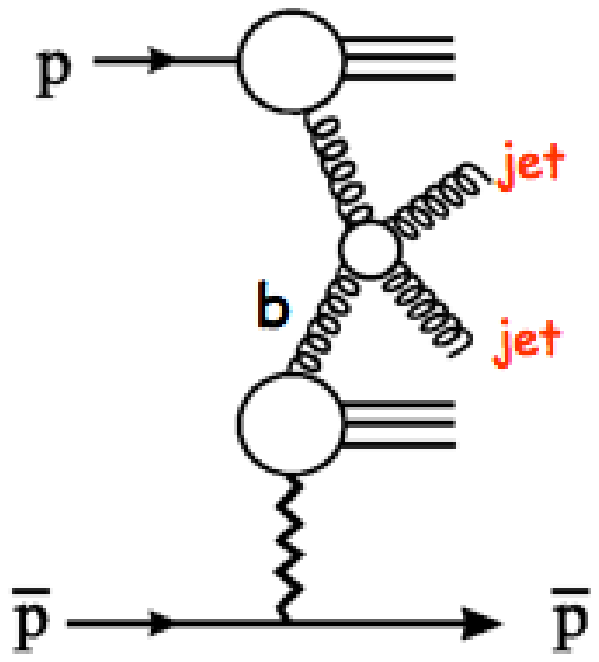} \hspace*{0.5cm}
\includegraphics[height=0.15\textheight]{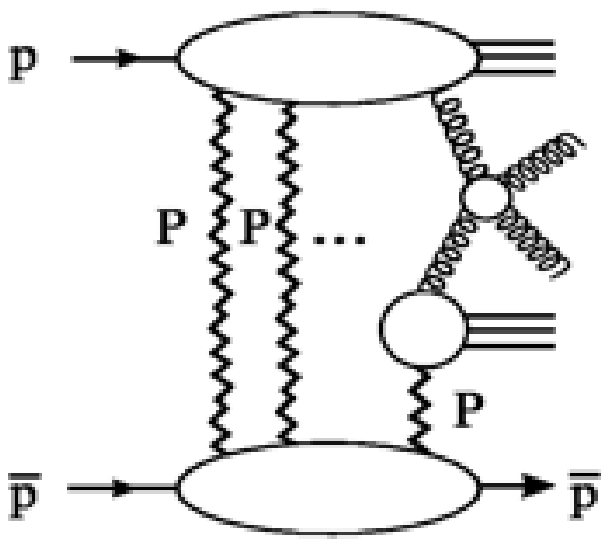}
\end{center}
\caption{Left, diagram for diffractive dijet production in $p\bar{p}$
  scattering. Right, diagram for diffractive dijet production with
  rescattering.}
\label{fig:diagrams}
\end{figure}

\begin{figure}[hbt]
\begin{center}
\includegraphics[width=0.4\textwidth]{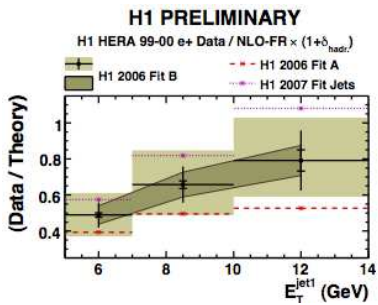}
\end{center}
\caption{The ratio of dijet diffractive cross section to NLO
  expectations obtained with various DPDFs assuming the validity of
  QCD factorization as a function of the highest transverse energy of
  the two jets, $\ETJ1$.  }
\label{fig:survival}
\end{figure}

QCD factorization in diffractive scattering is expected to break down
in hadron-hadron interactions~\cite{Collins:1998}. Indeed, the
measurements of diffractive dijet production in $p\bar{p}$
interactions at the Tevatron~\cite{cdfdiff} indicate that their rate is
by factor 5 to 10 lower than expected from the DPDF
extracted at HERA. This is shown in Fig.~\ref{fig:cdfdiff}, where
the relative rate of diffractively produced dijets is shown as a
function of $\beta$.
The suppression of hard diffractive scattering in $p\bar{p}$ is
understood as due to rescattering of partons which did not directly
participate in the diffractive process. As a result of the
rescattering, the LRG is destroyed. This is depicted schematically in
Fig.~\ref{fig:diagrams}.
Some suppression of LRG events is also expected in photoproduction
($\gp$) at HERA~\cite{klasen}, especially in the regime of resolved
photon contributions. In that respect, the data is not conclusive as
the H1 experiment observes a large suppression for both the resolved
photon (more hadron-like) and the direct photon (more DIS-like)
contributions~\cite{h1photo} while the ZEUS experiment sees very
little suppression if at all~\cite{zeusphoto}. The explanation may lie
in the preliminary measurements of the H1 experiment~\cite{h1prelim}
where the cross section for diffractive dijet production in $\gp$ is
compared to expectations (assuming QCD factorization) as a function of
the highest transverse energy of the two jets, $\ETJ1$. As shown in
Fig.~\ref{fig:survival}, the suppression of LRG events is strongest at
low $\ETJ1$.
This is an interesting observation which indicates that the selection
of higher $E_T$ jets renders rescattering less probable. In fact, a
closer inspection of the CDF results (see Fig.~\ref{fig:cdfdiff})
leads to a similar conclusion. The LRG suppression seems stronger at
high $\beta$, that is lower diffractive masses and therefore lower
$E_T$ jets. The selection of higher $E_T$ jets may be a way of
selecting a (transversely) smaller partonic configuration of the
dissociating particle, for which by virtue of color screening the
probability of rescattering becomes smaller. 

The comparison of the diffractive measurements at HERA and at the
Tevatron leads to the following puzzle. Of the order of 10\% of events
observed in $ep$ collisions cannot be accounted for in $p\bar{p}$
collisions, yet inclusive factorization seems to be preserved
(universality of proton PDF). The only explanation is that the
mechanism of rescattering somehow preserves the total cross section.
This must have implications for multi-parton interactions, one of the
big unknowns at LHC.

\section{Hard diffraction and the 3D structure of the proton}

\begin{figure}[hbt]
\begin{center}
\includegraphics[width=0.6\textwidth]{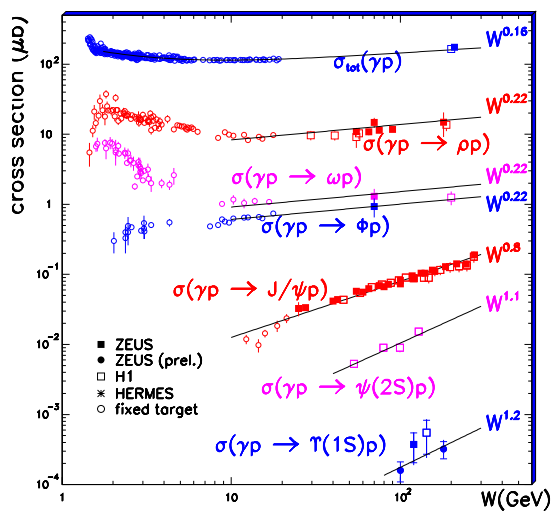}
\end{center}
\caption{Total and exclusive vector meson photoproduction cross sections, as a function of $W$. The curves are fits of the form $\sim W^\delta$ }
\label{fig:sigVM}
\end{figure}

A small fraction of diffractive events in $ep$ scattering at HERA
consist of exclusive vector meson or single photon (deeply virtual
Compton scattering DVCS) production. At the high center of mass
energies of HERA and in the presence of a large scale, these exclusive
processes are believed to be mediated by two gluon
exchange~\cite{ryskin,brodsky}.  The cross section for the exclusive
processes is expected to rise with center of mass energy $W$, with the
rate of growth increasing with the value of the hard scale. This is
illustrated in Fig.~\ref{fig:sigVM} where the $W$ dependence of the
$\gp$ cross section for exclusive $\rho$, $\phi$, $J/\psi$, $\psi(2S)$
and $\Upsilon$ production is shown together with the $W$ dependence of
the total $\gp$ cross section (for a detailed discussion and
references see~\cite{levy}). The $W$ dependence of the latter is
typical of soft interactions. Also shown in the figure are the values
of the logarithmic derivative $\delta$ obtained by fitting a
$W^\delta$ dependence.
Starting from exclusive $J/\psi$ production there is a definite change
in the $W$ dependence, with the tendency for $\delta$ to become larger
as the mass of the vector meson increases. This is interpreted as the
evidence of gluon participation in the interaction, since the higher the
scale at which the gluons are probed the faster the rise with $W$.

\begin{figure}[htbp]
\begin{minipage}{0.47\hsize}
\centerline{\includegraphics[width=\hsize]{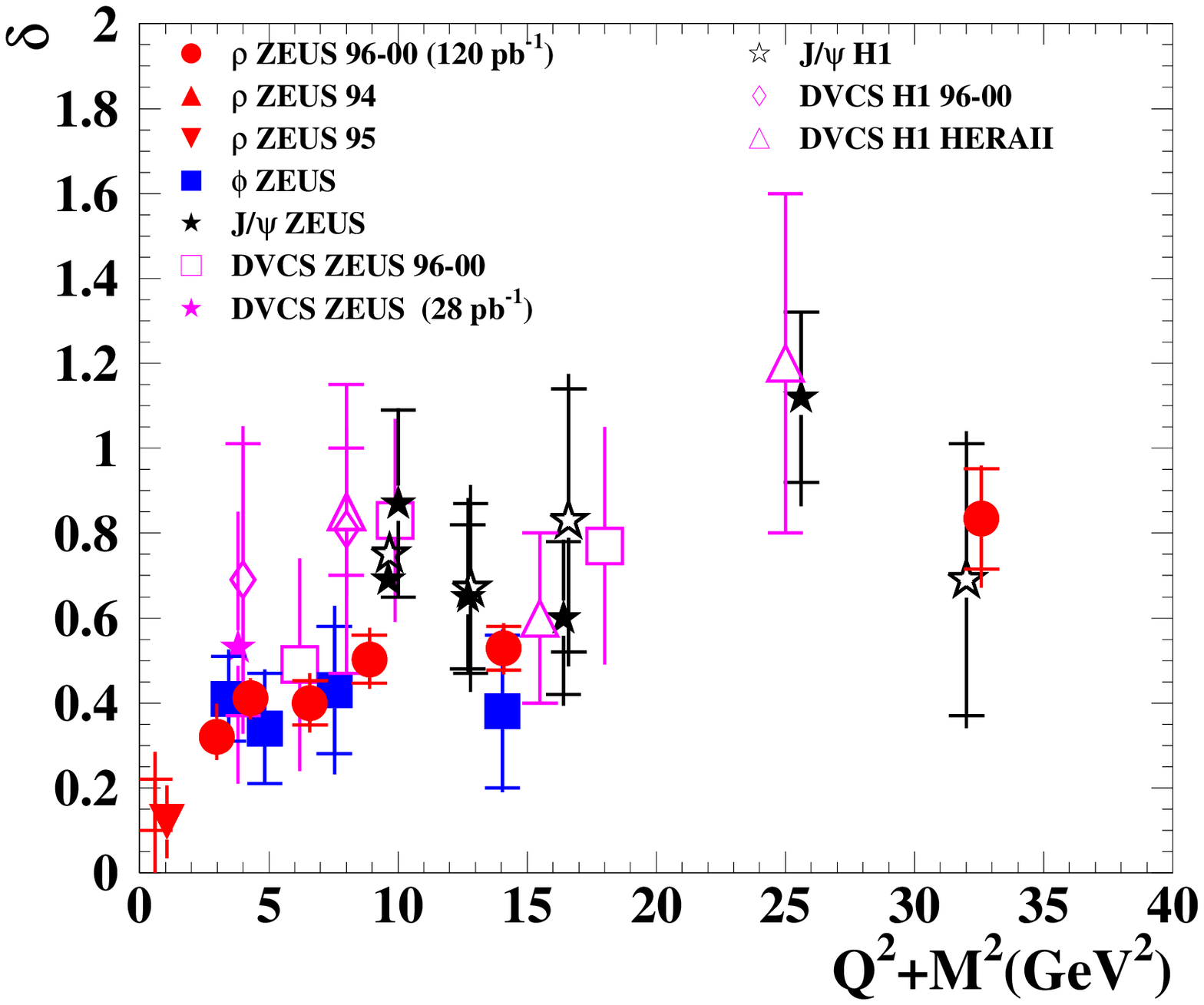}}
\end{minipage}
\begin{minipage}{0.47\hsize}
\centerline{\includegraphics[width=0.85\hsize]{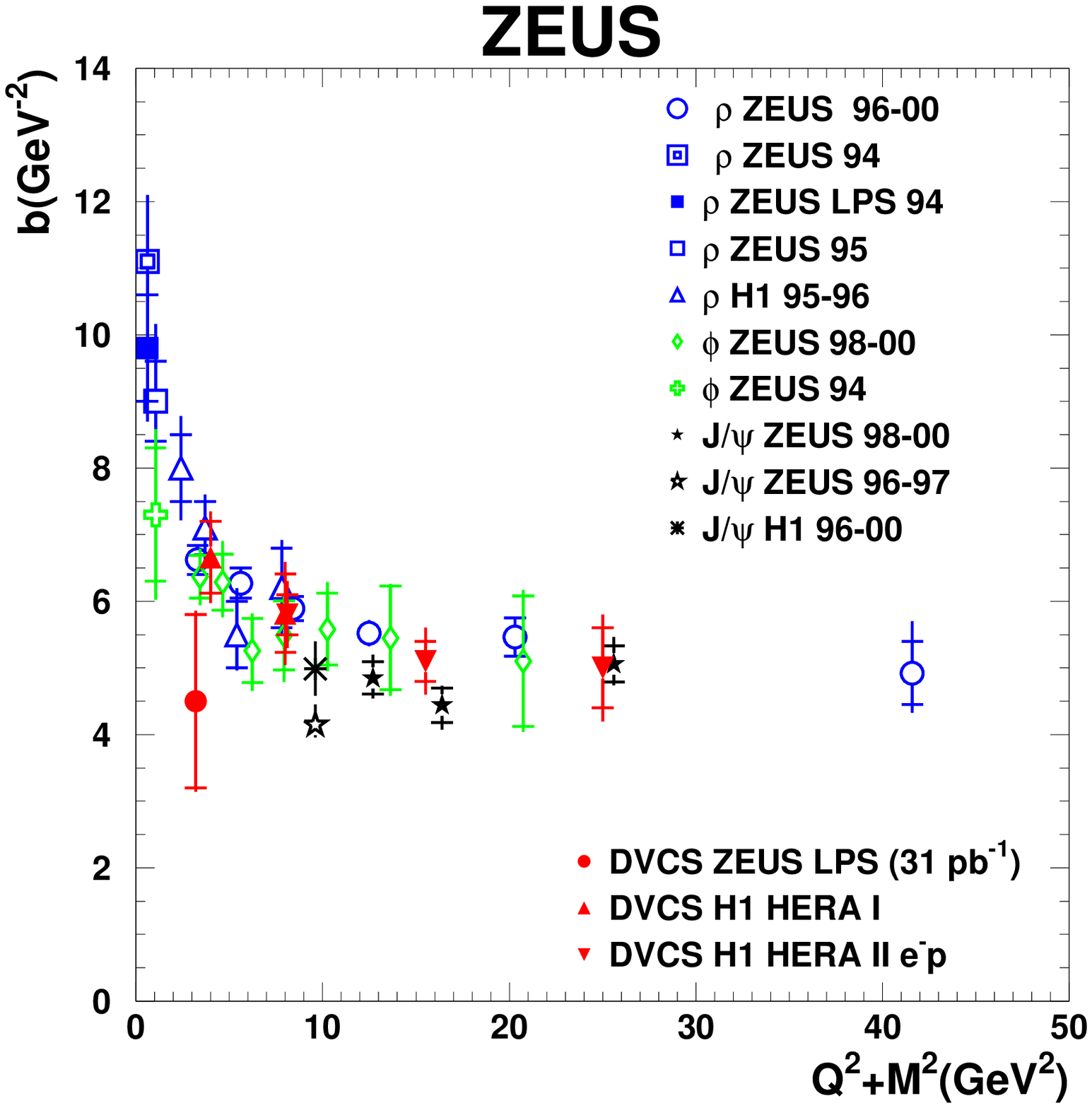}}
\end{minipage}
\caption{\it Left: logarithmic derivatives $\delta=d\log\sigma(\gvp)/d
  \log W$ as a function of $Q^2+M^2$ for exclusive processes.
  Right:Exponential slope of the $t$ distribution measured for
  exclusive processes as a function of $Q^2+M^2$.}
\label{fig:delta}
\end{figure}

A further compilation of logarithmic derivatives~\cite{levy}, which
also includes the measurements of exclusive processes in DIS, for
$\rho$, $\phi$, $J/\psi$ and DVCS as a function of scale defined as
$Q^2+M_{V}^2$, where $M_{V}$ is the mass of the vector meson, is
presented in figure~\ref{fig:delta}.
With increasing hardness of the interaction, the $t$ distribution is
expected to become universal, independent of the scale and of the
final state. This is because the harder the scale the more point-like
the virtual photon becomes. The exponential slope of the $t$
distribution, $b$, reflects then the size of the proton. A compilation
of measured $b$ values (see~\cite{zeus-dvcs} for details and
references therein) is presented in figure~\ref{fig:delta}.  Around
$Q^2+M^2$ of about $15 \gevtwo$ the $b$ values becomes universal.

Since exclusive vector meson production or DVCS at HERA energies is
driven by gluons, the $b$ value probes the size of the gluon cloud in
the proton~\cite{weiss}. The universal value of $b\simeq 5
\gevmtwo$ corresponds roughly to a radius of 0.6 fm which is
smaller than the 'charge' radius of the proton which is 0.8 fm.
This is the first indication that the gluons are well contained within
the proton. This finding has implications for multi-parton
interactions. In particular the more central the $pp$ interaction the
higher the probability of a second partonic interaction~\cite{weiss}.

\section{Hadronic final states}

At LHC, much of the background estimates rely on proper simulations of
the hadronic final states, of which multi-parton interactions are only
one of the elements. 

\begin{figure}[htbp]
\centerline{\includegraphics[width=0.9\hsize]{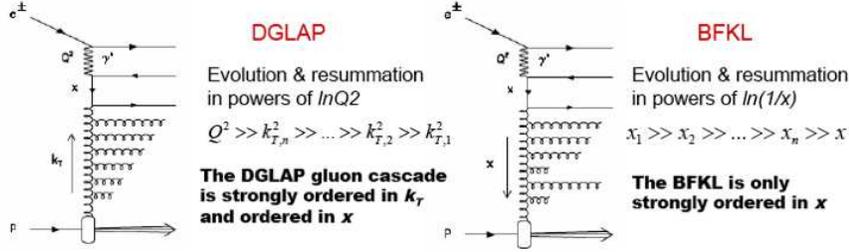}}
\caption{Schematic representation of the DGLAP (left) and BFKL (right)
  parton radiation patterns}
\label{fig:isrfsr}
\end{figure}

Most of the MC generators base the formation of the hadronic final
sate on the properties of the DGLAP dynamics, for which subsequent
emissions of partons, before and after the hard interaction, are
strongly ordered in transverse momentum $k_T$, with $k_T$ increasing
steadily while the momentum fraction $x$ decreases. At low $x$, where
effects of the BFKL~\cite{bfkl} dynamics may be felt, the ordering in
$k_T$ may be lost while subsequent emissions are strongly ordered in
$x$. The two different mechanisms are schematically depicted for DIS
in Fig.~\ref{fig:isrfsr}. Note that the contribution of large higher
order corrections to the DGLAP evolution may also appear as loss of
$k_T$ ordering.
The loss of strong $k_T$ ordering may result in the appearance of
high $E_T$ forward (in the proton direction) jets, of jets with
$E_T>Q$ and a decorrelation in the azimuthal angle between jets. All
of these effects have been observed at HERA.

\begin{figure}[htbp]
\centerline{\includegraphics[width=0.3\hsize]{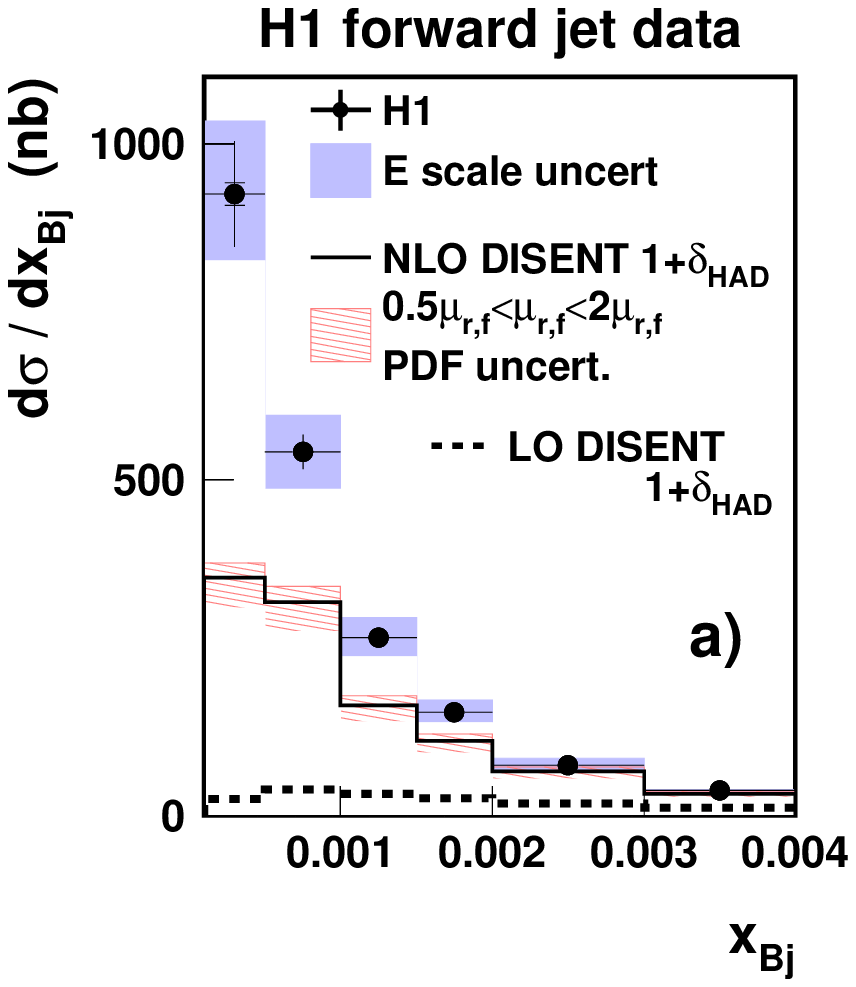}\includegraphics[width=0.3\hsize]{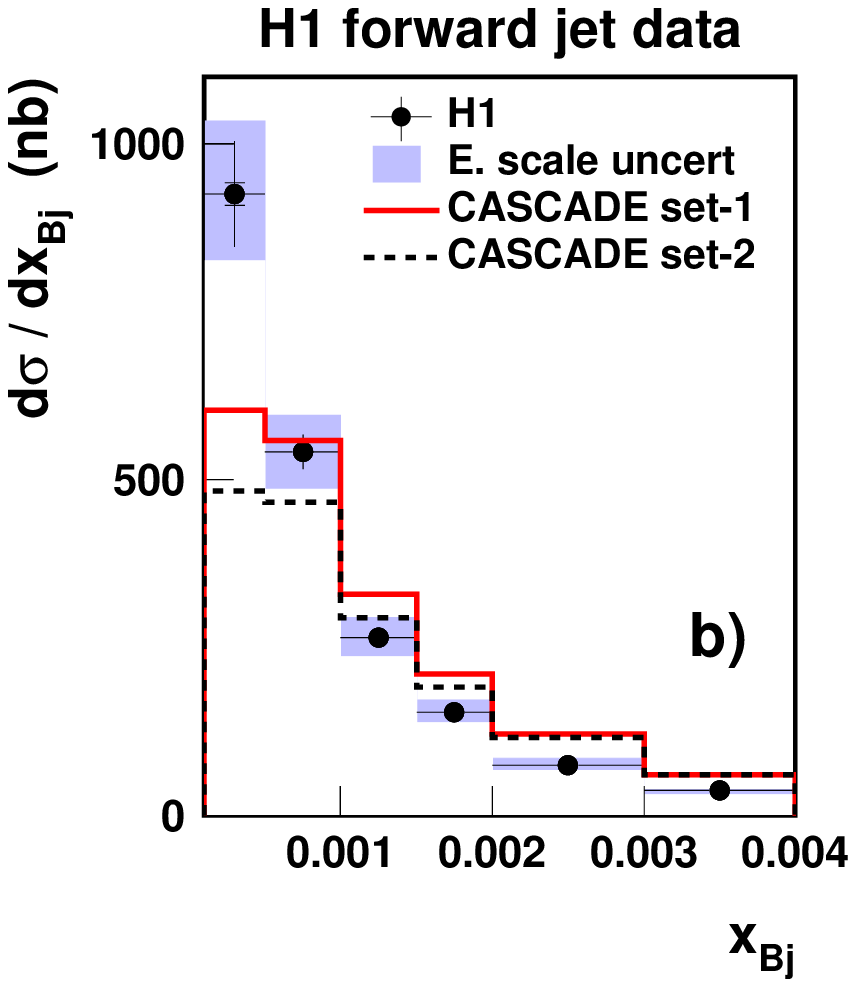}\includegraphics[width=0.3\hsize]{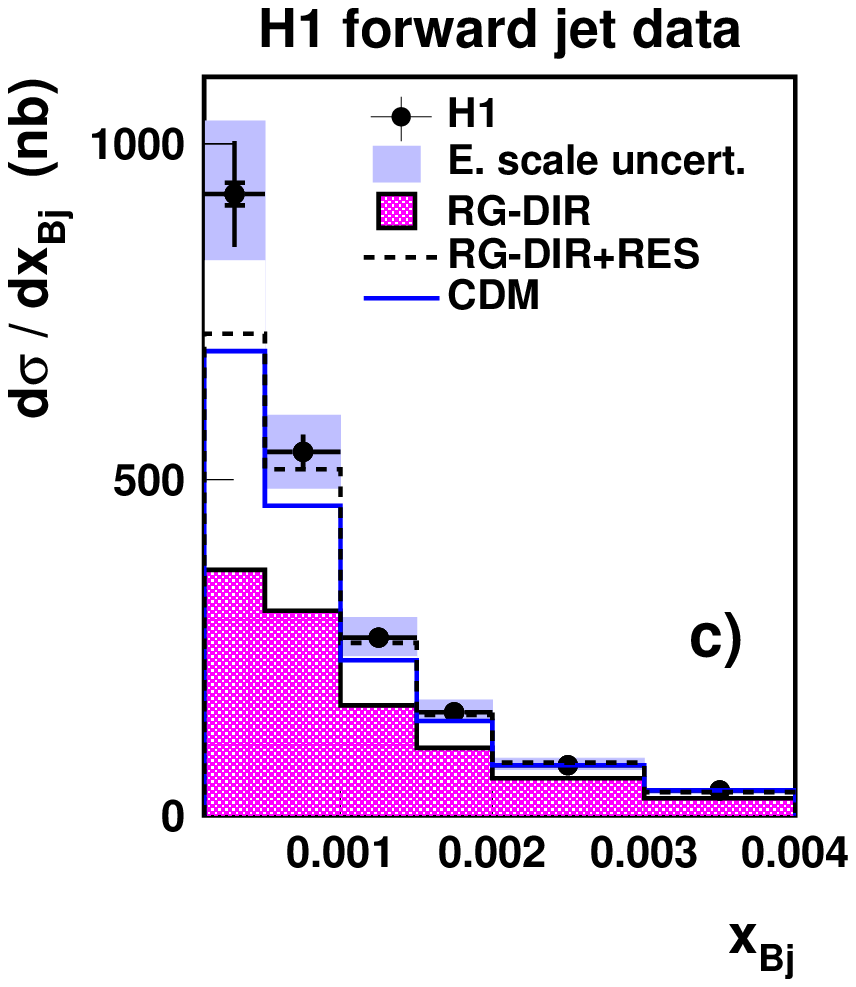}}
\caption{Cross section for producing forward jets, with $1.7<\ETAJ
  <2.8$, $x_\mathrm{jet} > 0.035$ and with transverse momentum $p_T$
  fulfilling $0.5<p_T^2/Q^2<5$ as a function of Bjorken $x$,
  $x_{\mathrm{Bj}}$. Also shown are expectations of NLO calculations
  and various MC generators.}
\label{fig:fjets}
\end{figure}

The cross section for producing forward jets, defined by $1.7<\ETAJ
<2.8$, $x_\mathrm{jet} > 0.035$ and with transverse momentum $p_T$
fulfilling $0.5<p_T^2/Q^2<5$, as measured by the H1
experiment~\cite{h1fjets}, is shown in Fig.~\ref{fig:fjets} as a
function of $x$. 
The measurements are compared to various MC generators. A clear excess
of forward jets is observed at low $x$, which cannot be accounted for
neither by higher order DGLAP corrections nor by any other MC
generator. The highest rate is expected in the Color Dipole Model
(CDM)~\cite{ariadne} generator in which the lack of $k_T$ ordering is
implicit in the way the color dipoles emit radiation.
\begin{figure}[htb]
\centerline{\includegraphics[width=0.65\hsize]{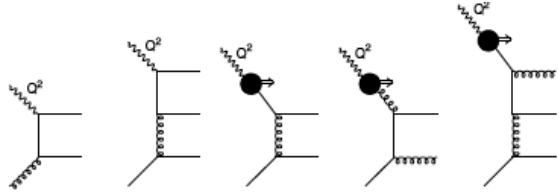}}
\caption{Schematic representation of diagrams with direct (DIS) and
  resolved $\gv$ contributions included in the simulation of DIS
  hadronic final states.}
\label{fig:resgv}
\end{figure}
If the DGLAP dynamics, as represented by the RG
generator~\cite{rapgap}, is supplemented by the addition of a resolved
virtual photon contribution~\cite{jung}, the rate of forward jet
production increases but still fails to describe the lowest $x$
measurements.  The addition of a resolved virtual photon contribution
is supposed to mimic the lack of $k_T$ ordering as shown schematically
in Fig.~\ref{fig:resgv}.

\begin{figure}[htb]
\centerline{\includegraphics[height=0.40\textheight]{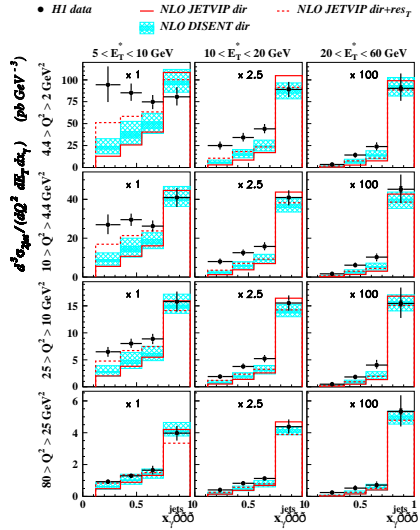}}
\caption{Cross section for dijet production as a function of
  $x_\gamma^{\mathrm{jets}}$, the estimator of the $\gv$ momentum
  carried by the two jets, for various ranges of transverse energy
  measured in the $\gvp$ center of mass system, $E_T^*$, and $Q^2$. Also
  shown are predictions of NLO calculations with and without the
  resolved $\gv$ contribution.}
\label{fig:resjets}
\end{figure}The addition of a resolved photon contribution also helps in
describing the cross section for dijet production in DIS, in
particular in the regime where the $E_T^2$ of the jets is higher than
$Q^2$~\cite{resdis}. This is shown in Fig.~\ref{fig:resjets} where the
dijet differential cross section is plotted as a function of
$x_\gamma^{\mathrm{jets}}$, the estimator of the $\gv$ momentum
carried by the two jets, for various ranges of transverse energy
measured in the $\gvp$ center of mass system, $E_T^*$, and $Q^2$.
However even the addition of the resolved photon contribution falls
short of describing the low $Q^2$ region where $x$ is lowest.

\begin{figure}[htbp]
\begin{center}
 \centerline{\includegraphics[width=0.45\hsize]{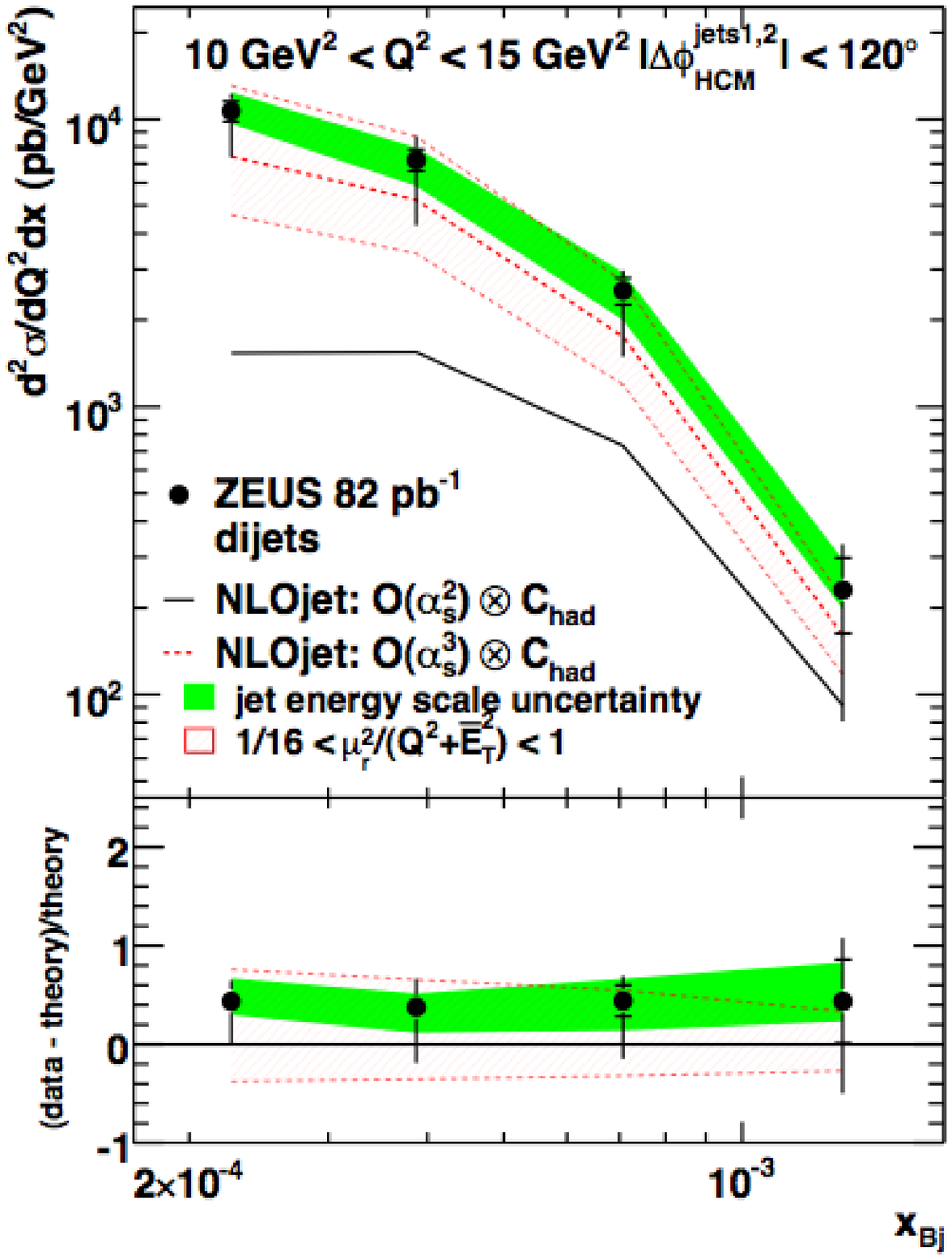}\hfill \includegraphics[width=0.45\hsize]{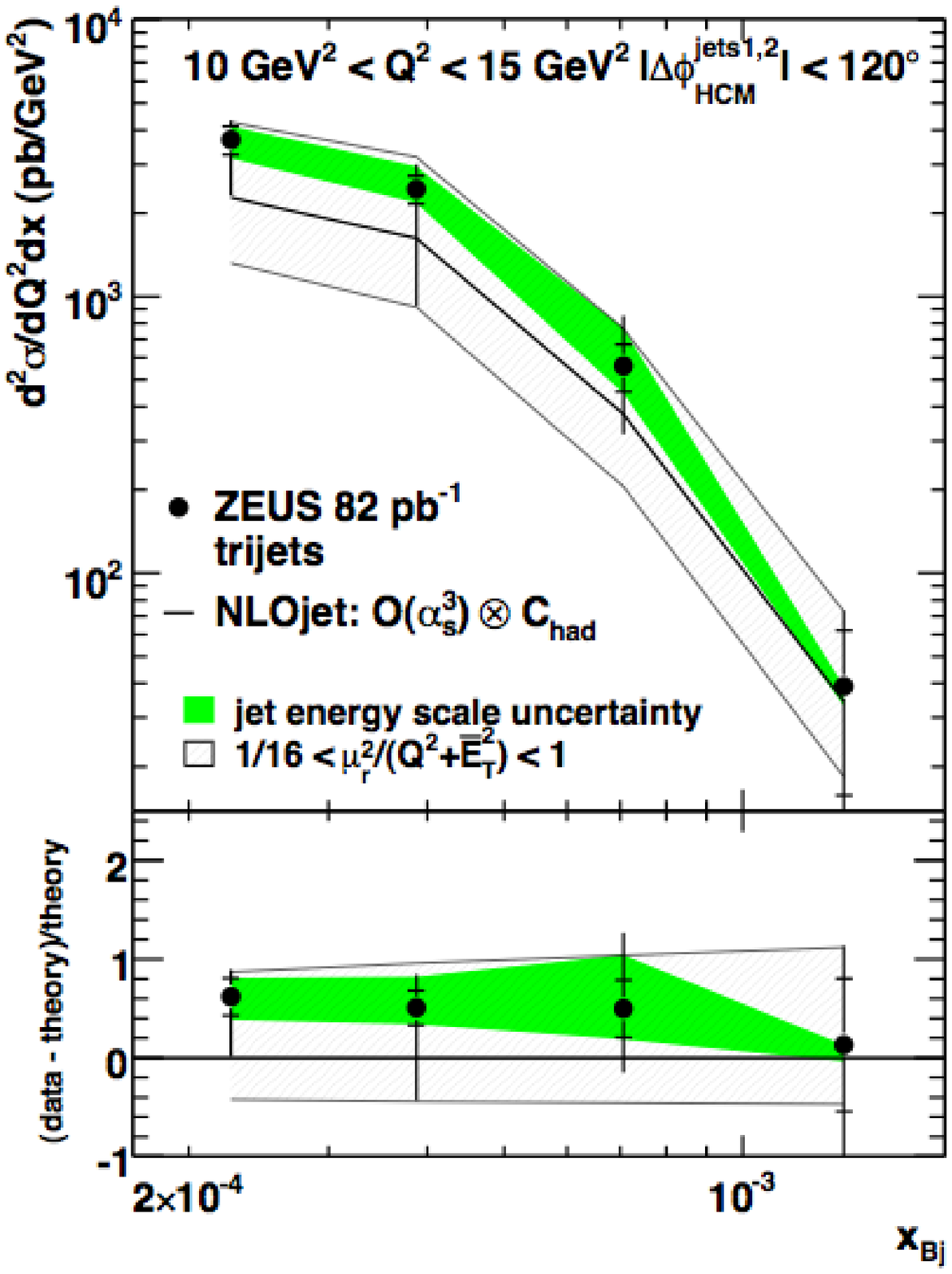}}
\end{center}
  \caption{DIS cross section for dijet and trijet production, when the
    two highest $E_T$ jets are separated by an azimuthal angle
    $\Delta \phi<120^\circ$, as a function of Bjorken $x$ -
    $x_{\mathrm{Bj}}$. Also shown are the expectations of NLO
    calculations without and with  ${\cal{O}}(\alpha_s^3)$ corrections.}
\label{fig:deltaphi}
\end{figure}

The azimuthal correlations in multijet production have been
measured both by the H1~\cite{h1multijet} and the
ZEUS~\cite{zeusmultijet} experiments. The NLO calculations of dijet
production fail to describe the rate of jets with the difference
in azimuthal angle $\Delta \phi<120^\circ$ as shown in
Fig.~\ref{fig:deltaphi}, in particular at low $x$. However, an addition
of $\alpha^3$ terms seems to cure the problem and also properly
describes the rates for trijet events.

All these findings strongly suggest that at low $x$ the description of
hadronic final states requires corrections even beyond NLO DGLAP.

\section{Summary}

The large fraction of DIS diffractive events, mainly driven by gluons,
could suggest the approach to the unitarity limit, in which case
non-linear effects could bias the measurements. The observation of an
excess of forward jets in the low $x$ regime would suggest that the
transverse momentum ordering of gluon radiation, inherent in the DGLAP
evolution, breaks down at low $x$.  The same conclusion may be derived
from the appearance of the so called 'resolved photon' contribution
necessary to described DIS dijet production in the low $x$ regime. At
HERA, all these effects constitute a relatively small fraction of the
total cross section (typically 5\%) and therefore may not be visible
in the fully inclusive measurements.  At LHC, these low $x$ effects may
be greatly enhanced. In particular, the increased jet multiplicity and
the shift of the high transverse momentum activity toward the proton
direction could result in events with missing transverse energy
greater than expected from Monte Carlo generators.

The non-uniform space distribution of partons in the proton, partly
covered by the newest generation of MC programs, and the strong
breaking of diffractive QCD factorization due to rescattering, with
potentially non-trivial dependence on kinematics, may lead to
surprises in modeling the underlying event and multi-parton
interactions accompanying hard $pp$ interactions at the LHC.

\section{Acknowledgements}
I would like to thank the organizers of the meeting for inviting me to
this stimulating event in memory of Prof. Jan Kwiecinski. I myself
belong to the generation of physicists who were inspired by Jan and
much of our perseverance at HERA in exploring the QCD dynamics at low
x is due to his teachings.

\end{document}